\documentclass[twocolumn,showpacs,preprintnumbers,amsmath,amssymb,aps,prc,superscriptaddress,reprint,floatfix]{revtex4-2}

\usepackage{graphicx}
\usepackage{dcolumn}
\usepackage{bm}

\usepackage{mathtools}
\usepackage{amsmath}
\usepackage{mathrsfs}
\usepackage{hyperref}
\usepackage{cleveref}[2012/02/15]
\crefformat{footnote}{#2\footnotemark[#1]#3}
\usepackage{lineno}
\usepackage{xcolor}
\usepackage[normalem]{ulem}

\begin{document}


\title{Extraction of neutron-capture cross sections on $^{92}$Zr using the charge-exchange Oslo method}

\author{N.D. Pathirana}
\email{devapath@frib.msu.edu}
\affiliation{Facility for Rare Isotope Beams, Michigan State University, East Lansing, MI 48824, USA\looseness=-1}
\affiliation{Department of Physics and Astronomy, Michigan State University, East Lansing, MI 48824, USA\looseness=-1}

\author{R.G.T. Zegers}
\email{zegers@frib.msu.edu}
\affiliation{Facility for Rare Isotope Beams, Michigan State University, East Lansing, MI 48824, USA\looseness=-1}
\affiliation{Department of Physics and Astronomy, Michigan State University, East Lansing, MI 48824, USA\looseness=-1}

\author{B. Gao}
\affiliation{State Key Laboratory of Heavy Ion Science and Technology, Institute of Modern Physics, Chinese Academy of Sciences, Lanzhou 730000, China\looseness=-1}

\author{A. Spyrou }
\affiliation{Facility for Rare Isotope Beams, Michigan State University, East Lansing, MI 48824, USA\looseness=-1}
\affiliation{Department of Physics and Astronomy, Michigan State University, East Lansing, MI 48824, USA\looseness=-1}

\author{A.C. Larsen}
\affiliation{Department of Physics, University of Oslo, NO-0316 Oslo, Norway\looseness=-1}

\author{H. Berg}
\affiliation{Facility for Rare Isotope Beams, Michigan State University, East Lansing, MI 48824, USA\looseness=-1}
\affiliation{Department of Physics and Astronomy, Michigan State University, East Lansing, MI 48824, USA\looseness=-1}

\author{D. Bazin}
\affiliation{Facility for Rare Isotope Beams, Michigan State University, East Lansing, MI 48824, USA\looseness=-1}
\affiliation{Department of Physics and Astronomy, Michigan State University, East Lansing, MI 48824, USA\looseness=-1}

\author{H.L. Crawford}
\affiliation{Lawrence Berkeley National Laboratory, Berkeley, CA 94720, USA\looseness=-1}

\author{A. Gade}
\affiliation{Facility for Rare Isotope Beams, Michigan State University, East Lansing, MI 48824, USA\looseness=-1}
\affiliation{Department of Physics and Astronomy, Michigan State University, East Lansing, MI 48824, USA\looseness=-1}

\author{P. Gastis}
\affiliation{Los Alamos National Laboratory, Los Alamos, NM 87545, USA\looseness=-1}

\author{T. Ginter}
\affiliation{Facility for Rare Isotope Beams, Michigan State University, East Lansing, MI 48824, USA\looseness=-1}

\author{C.J. Guess}
\affiliation{Department of Physics and Astronomy, Rowan University, Glassboro, NJ 08028, USA\looseness=-1}

\author{M. Guttormsen}
\affiliation{Department of Physics, University of Oslo, NO-0316 Oslo, Norway\looseness=-1}

\author{S. Noji}
\affiliation{Facility for Rare Isotope Beams, Michigan State University, East Lansing, MI 48824, USA\looseness=-1}

\author{B. Longfellow}
\affiliation{Facility for Rare Isotope Beams, Michigan State University, East Lansing, MI 48824, USA\looseness=-1}
\affiliation{Department of Physics and Astronomy, Michigan State University, East Lansing, MI 48824, USA\looseness=-1}

\author{J. Pereira}
\affiliation{Facility for Rare Isotope Beams, Michigan State University, East Lansing, MI 48824, USA\looseness=-1}

\author{L.A. Riley}
\affiliation{Department of Physics and Astronomy, Ursinus College, Collegeville, Pennsylvania 19426, USA\looseness=-1}

\author{D. Weisshaar}
\affiliation{Facility for Rare Isotope Beams, Michigan State University, East Lansing, MI 48824, USA\looseness=-1}

\author{J.C. Zamora}
\affiliation{Facility for Rare Isotope Beams, Michigan State University, East Lansing, MI 48824, USA\looseness=-1}

\date{\today}

\begin{abstract}
The $^{93}$Nb($t$,$^{3}$He) reaction at 115 MeV/u was studied to demonstrate that nuclear level densities and $\gamma$-ray strength functions can be extracted from charge-exchange reactions at intermediate energies using the Oslo technique. The matrix of excitation energy in $^{93}$Zr, reconstructed from the ($t$,$^{3}$He) reaction, versus the energy of $\gamma$ rays emitted by the excited $^{93}$Zr nuclei, was obtained in an experiment with the S800 Spectrograph operated in coincidence with the GRETINA $\gamma$-ray detector. The extracted level density and $\gamma$-ray strength function obtained by applying the Oslo method to this matrix were used to estimate the $^{92}$Zr($n$,$\gamma$)$^{93}$Zr cross section by combining the new results with other experimental data and theoretical calculations for $E$1 and $M$1 strength functions at higher energies. Good agreement with direct measurements of the $^{92}$Zr($n$,$\gamma$)$^{93}$Zr cross section was found. The contribution from the upbend in the extracted $\gamma$-ray strength function was important to achieve the consistency as the neutron-capture cross section without this contribution is significantly below the direct measurements otherwise. Since charge-exchange reactions at intermediate energies have long been used for extracting Gamow-Teller strengths, the successful demonstration of the charge-exchange Oslo method enables experiments in which ($n$,$\gamma$) cross sections and Gamow-Teller strengths can be measured simultaneously, which is of benefit for astrophysical studies.   

\end{abstract}

\maketitle

                             
\maketitle

\section{\label{sec:level1}Introduction}
Neutron captures play an important role in a variety of topics. The nucleosynthesis of elements beyond iron primarily takes place in astrophysical environments in which the density of neutrons is high \cite{meyer1994,RevModPhys.29.547}. These environments allow for the slow ($s$), \cite{pignatari2010,F.Kappeler2011}, fast ($r$) \cite{Arnould2007}, and intermediate ($i$) \cite{COW77} neutron-capture processes to occur. The type of process in a given astrophysical environment depends on the competition between temperature-dependent neutron-capture reactions, which make the abundances more neutron-rich, and $\beta$ decay, which pushes the abundances towards the valley of stability \cite{SNE08}. Neutron captures also play an important role in production rates of isotopes that can act as cosmic chronometers \cite{CLA64,PhysRevC.82.015802,travaglio2014}. Although this work is motivated by the astrophysical applications, neutron-capture reactions are also important for the design of nuclear power generators \cite{ALIBERTI2006700,C0EE00108B}. An accurate knowledge of neutron-capture rates is necessary for understanding the best conditions for the burning of nuclear waste \cite{C0EE00108B,etde_22491258}, which, in turn, is important for managing long-term storage of this waste. Finally, neutron captures play an impactful role in other applications, such as the production of isotopes and neutron-activation analyses \cite{rubbia1997resonance, ABDELNOUR2025112693}.
\setlength{\parskip}{0.25em}

Although direct measurements of neutron-capture cross sections on nuclei that are stable or long-lived are feasible, this is not the case for nuclei that are unstable and have short half-lives, as target foils cannot be produced and a neutron target is not available. Since several of the applications, in particular for understanding astrophysical phenomena, involve many short-lived isotopes, one has to rely on theory or indirect experimental methods for estimating the neutron-capture cross sections. The use of theoretical methods to estimate neutron capture cross sections leads to significant uncertainties \cite{PhysRevC.64.035801,Liddick2016}. Therefore, indirect experimental methods are highly desirable to constrain the theoretical uncertainty band. Indirect measurements provide neutron-capture cross sections for specific nuclei and can guide the development of improved theoretical models that can be used for a larger set of nuclei \cite{LARSEN2019}. Techniques that have been developed for the indirect study of neutron-capture cross section include the $\gamma$-ray strength method \cite{PhysRevC.82.064610}, the surrogate reaction method \cite{RevModPhys.84.353, PhysRevLett.121.052501}, the Oslo method \cite{GUTTORMSEN1987, SCHILLER2000, LARSEN2019}, the inverse-Oslo method \cite{Ingeberg2020,Ingeberg2025}, and derived from the latter, the $\beta$-Oslo method \cite{Artemis2014, Liddick2016, LARSEN2019}. 

The Oslo method is used to extract the nuclear level density (NLD) and $\gamma$-ray strength function ($\gamma$SF) from an experimentally obtained matrix of excitation energy versus $\gamma$\-ray energies. To excite the nucleus of interest, different nuclear reactions have been employed \cite{LARSEN2019}. In the $\beta$-Oslo method, $\beta$ decay is used to populate the nucleus of interest. A statistical model for the neutron-capture reaction is used to estimate the cross section. The NLD provides information about the nuclear energy levels available in the nucleus produced by neutron capture and the $\gamma$SF provides information about the decay properties of the levels reached in the ($n$,$\gamma$) reaction. To estimate the neutron-capture cross section, the interaction between the neutron and the capturing nucleus must also be modeled, which is done using the optical model potential. 

In this work, we introduce the charge-exchange (CE) Oslo method. In this method, CE reactions at intermediate energies of $\gtrsim 100$ MeV/u are used to populate the nucleus of interest, adding another type of reaction to those already used for the Oslo method. An added benefit of using CE reactions at intermediate energies is that they also provide a tool for the model-independent extraction of Gamow-Teller (GT) transition strengths, including transitions with $Q$ values inaccessible through electron capture (EC) or $\beta$ decay studies. The extraction of GT strength is based on the proportionality between GT strength and the CE cross section at small momentum transfer ($q\approx0$) \cite{Tad87,Zeg06}. The extraction of GT strength up to high excitation energies is important for constraining weak interaction rates in astrophysical phenomena \cite{RevModPhys.75.819,stars2021}. Hence, by using the CE-Oslo method, the extraction of neutron-capture cross sections and weak-interaction rates can be performed in a single measurement. Here, we perform the CE-Oslo based on an experiment performed with a light-ion beam impinging on a stable target. In the future, the goal is to extend the method to ($p$,$n+\gamma$) experiments at intermediate beam energies performed in inverse kinematics aimed at studying unstable nuclei, by combining ($p$,$n$) experiments in inverse kinematics (see e.g., \cite{Sas11,PhysRevLett.121.132501}) with $\gamma$ detection.       

The effectiveness of the CE-Oslo method is demonstrated using the $^{93}$Nb($t$,$^{3}$He+$\gamma$) reaction. This reaction was previously studied for the extraction of the GT strength distribution in the $\beta^{+}$ direction for $^{93}$Nb \cite{Gao2020}. The same data are now used for the indirect determination of the $^{92}$Zr($n,\gamma$) cross section. The latter reaction has been studied directly \cite{BOLDEMAN1976,MOSKALEV1964667,macklin1967,PhysRev.165.1329,Bartholomew1973, OHGAMA2005, Tagliente2010,Tagliente2022}, which is useful to test the indirect extraction of the neutron-capture cross section via the CE-Oslo method.  In Refs. \cite{OHGAMA2005,macklin1967}, the $^{92}$Zr($n,\gamma$)$^{93}$Zr cross sections were measured. In Refs. \cite{macklin1967,BOLDEMAN1976,Bao2000, Tagliente2010,Tagliente2022}, Maxwellian averaged cross sections (MACS) for the $^{92}$Zr($n,\gamma$)$^{93}$Zr reaction were determined. Here, the results from the CE-Oslo method are compared with both types of previously obtained cross sections. Whereas the extraction of MACS for $k_{B}T < 30$ keV, where $k_{B}$ is the Maxwell-Boltzmann constant and $T$ is the temperature, was previously found to be consistent across different measurements, the extraction of MACS for $k_{B}T > 30$ keV, which requires the use of capture cross section for relatively high-energy neutrons, was reported to be uncertain \cite{Tagliente2022}. The present results provide new information about MACS at higher temperatures. The neutron-capture rate on $^{93}$Zr is of interest as input to $s$ process simulations probing the origin of stardust \cite{Lugaro2014}.          

\begin{figure*}
\hspace*{0cm}
\centering
\includegraphics[width=1.01\textwidth,clip]{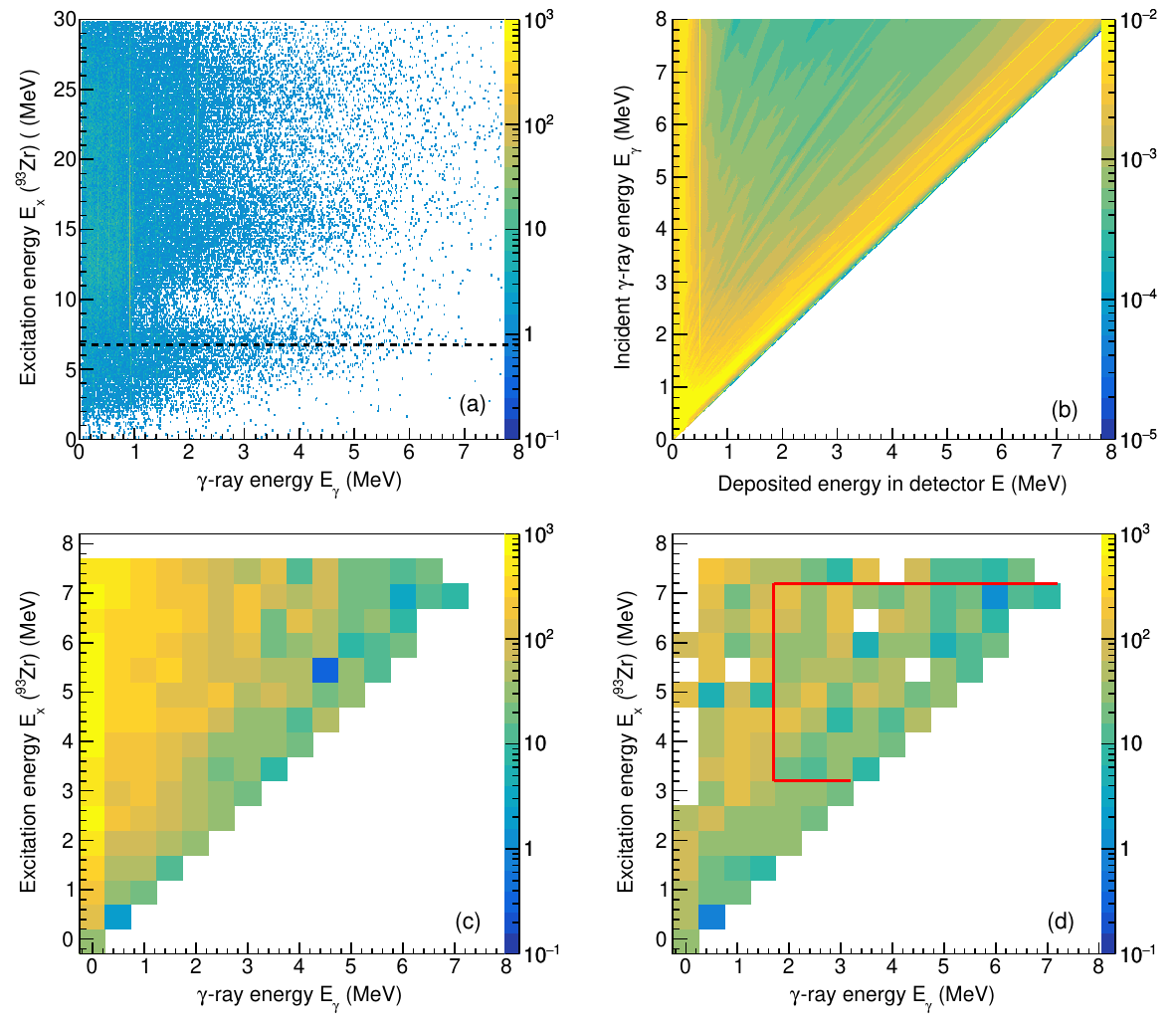}
\caption{(a) Experimental $E_{x}$--$E_{\gamma}$ particle-$\gamma$ coincidence raw matrix for the $^{93}$Nb($t$,$^{3}$He$+\gamma$) reaction. The x-axis bin size is 10 keV, while the y-axis bin size is 100 keV. The dashed black line represents the neutron separation energy in $^{93}$Zr. The excitation energy was measured up to $E_{x}(^{93}\mathrm{Zr}) = 30$ MeV as shown in the figure. However, only excitation energies up to 7.2 MeV were included in the CE-Oslo analysis; (b) the GRETINA response function obtained from the UCGRETINA simulation. The x-axis bin size is 10 keV, while the y-axis bin size is 10 keV; (c) the $E_{x}$--$E_{\gamma}$ matrix after unfolding of the detector response. The x-axis bin size is 500 keV, while the y-axis bin size is 500 keV; (d) primary $E_{x}$--$E_{\gamma}$ matrix, which is used in the CE-Oslo method. The x-axis bin size is 500 keV, while the y-axis bin size is 500 keV. The area within the solid red lines indicates the region used for the extraction of the NLD and the $\gamma$SF through the Oslo analysis. } 
\label{fig:matrices}
\end{figure*}

The structure of the paper is as follows. Section \ref{sec:level2} provides a brief overview of the $^{93}$Nb($t$,$^{3}$He+$\gamma$) experiment. In Section \ref{sec:level3}, the CE-Oslo method is explained, detailing the extraction of the primary $\gamma$-ray matrix, NLD, and $\gamma$SF for $^{93}$Zr. Section \ref{sec:level4} details the normalization of the NLD and the $\gamma$SF of $^{93}$Zr. In section \ref{sec:level5}, the $^{92}$Zr($n,\gamma$) cross sections and MACS are obtained and compared with cross sections previously measured. Finally, Section \ref{sec:level6} provides a summary and outlook.

\section{\label{sec:level2}Experiment}
The $^{93}$Nb($t$,$^{3}$He+$\gamma$) data used in the CE-Oslo analysis are the same as used for the results published previously in Ref. \cite{Gao2020}. In that reference, the focus was on the extraction of GT transition strength in the $\beta^{+}$ direction from $^{93}$Nb, populating final states in $^{93}$Zr. Here, we briefly restate the experimental details that are relevant for the CE-Oslo analysis. 

The experiment was performed at the Coupled Cyclotron Facility of the National Superconducting Cycltoron Laboratory (NSCL). By fragmenting a 150 MeV/u primary $^{16}$O beam with an intensity of 150 pnA, a secondary triton beam  at 115 MeV/u with a momentum spread of 0.5\% was produced with an intensity of 3$\times10^{6}$ pps and a purity in excess of 99\%. The beam was transported to the pivot point of the S800 Spectrograph \cite{Bazin2003}, where a $^{93}$Nb target foil with a thickness of 34 mg/cm$^{2}$ and a purity of 99.9\% was placed. Scattered $^{3}$He particles were identified and their momentum analyzed in the focal plane of the S800 Spectrograph \cite{YURKON1999291}. The dispersion of the beam on the target was matched to the dispersion of the S800 Spectrograph. Therefore, the momentum dispersion of in the incoming beam was canceled in the transport of scattered particle through the spectrograph \cite{FUJITA200217}. This allows for the determination of the excitation energy in $^{93}$Zr, which is obtained through a missing-mass calculation, with a resolution of 500 keV (FWHM), much smaller than the energy spread in the triton beam ($\sim$ 3.3 MeV). Note that the difference in energy loss in the $^{93}$Nb foil between a triton and an $^{3}$He particle was $\sim 400$ keV, limiting the energy resolution that could be achieved. 

The de-excitation $\gamma$-rays from $^{93}$Zr, or one of its decay products if the excitation energy exceeded the threshold for neutron or proton emission, were detected in the Gamma-Ray Energy Tracking In-beam Nuclear Array, GRETINA. \cite{PASCHALIS201344,WEISSHAAR2017187}. It was equipped with thirty-two 36-fold segmented high-purity germanium detectors, covering about 1$\pi$ sr of solid-angle. The photo-peak detection efficiency was approximately 4\% for a $\gamma$-ray energy $E_{\gamma}$ = 2 MeV. The $\gamma$ rays are produced approximately at rest and no Doppler reconstruction was required to determine their energies in the rest frame of the decaying nucleus. The energy resolution was about 0.2\% (FWHM) at 1332 keV.                

By combining the excitation energy ($E_{x}$) determined from the measurement of the $^{3}$He ejectile in the S800 Spectrograph with the $\gamma$-ray energies ($E_{\gamma}$) obtained from GRETINA, the ``raw'' experimental particle-$\gamma$ coincidence matrix was constructed, as shown in Fig. \ref{fig:matrices}(a). The raw coincidence matrix used in this paper was directly taken from the analysis described in Ref. \cite{Gao2020}. The coincidence matrix is almost background free, except near $E_{x}(^{93}$Zr)= 0 MeV, where minor contributions from ($t$,$^{3}$He) reactions on hydrogen that was absorbed on the $^{93}$Nb target foil led to the production of neutrons that can interact with GRETINA and produce background photons. This background is outside of the region of interest for the Oslo analysis and does not impact the further analysis.        

\section{\label{sec:level3}CE-Oslo Method}
As discussed above, the $E_{x}$--$E_{\gamma}$ matrix is shown in Fig. \ref{fig:matrices}(a). The excitation-energy spectrum was measured up to $E_{x}(^{93}$Zr)$=30$ MeV. This is much higher than the threshold for decay by particle emission. If neutrons and/or protons are emitted, a nucleus different from $^{93}$Zr is populated and the data cannot be used in the Oslo analysis. The neutron separation energy ($S_{n}$) of $^{93}$Zr is 6.734 MeV, indicated by the dashed black line in Fig. \ref{fig:matrices}(a). However, the drop in $\gamma$ yield, indicative of the channel for the decay by neutron emission opening, only occurs at around 8 MeV. As discussed in Ref. \cite{Gao2020}, this is due to the population of states with relatively high spin. Since the ground state of $^{93}$Nb has a spin-parity ($J^{\pi}$) of $9/2^{+}$ and in CE reactions near $q=0$ only a few units of angular momentum can be transferred ($\Delta L \lesssim 2$), states populated in $^{93}$Zr will have relatively high spins as well. Decay by neutron emission will, therefore, be hindered by the angular momentum barrier, delaying the onset of decay by neutron emission. As a result, excitation energies up to 7.2 MeV were included in the CE-Oslo analysis.

To obtain the NLD and $\gamma$SF from the experimental data using the Oslo method, four steps are required, which are described in this section. The analysis was performed with the Oslo Method Code Package \cite{magneg_2022}. 

In the first step, the raw $E_{x}$--$E_{\gamma}$ matrix of Fig. \ref{fig:matrices}(a) is unfolded to account for the $\gamma$-ray detector response. The unfolding method is described in Ref. \cite{GUTTORMSEN1996}. The response function of GRETINA, which describes the deposited energy distribution as a function of incident $\gamma$ energy, is shown in Fig. \ref{fig:matrices}(b). It was obtained using the UCGRETINA GEANT4 simulation \cite{RILEY2021}, which has been validated against data taken with calibration sources. 

As part of the Oslo analysis, one has to decide on the energy bin widths used, with the constraint that the excitation-energy bin width must be identical to the $\gamma$-energy bin width to ease the uncertainty analysis. In this work, the resolution in $E_{\gamma}$ is about 200 times smaller than that in $E_{x}$. Therefore, the energy bin width was chosen to match the latter, 0.5 MeV (FWHM). To test whether the choice of bin width impacted the results, analyses with narrower energy bin widths were also performed. No systematic differences in the results were observed, but the statistical uncertainties increased when narrow energy bin widths were used. Figure \ref{fig:matrices}(c) depicts the unfolded $\gamma$-ray spectrum using the energy binning of 0.5 MeV.

The next step in the Oslo analysis is to reduce the $E_{x}$--$E_{\gamma}$ matrix that has been unfolded with the detector response and which contains all $\gamma$ rays from the decay cascade, to a matrix that only contains the first-generation or primary $\gamma$ ray in the cascade. The main assumption underlying this procedure is that the $\gamma$-ray spectra are the same regardless of whether a state is populated directly by the CE reaction or through a $\gamma$ decay from a higher-lying state. To extract the NLD and the $\gamma$SF from the primary matrix, the excited nucleus must have fully equilibrated after the excitation, so that the decay process is statistical in nature. Minority non-statistical (i.e., direct or semi-direct) contributions from decay by particle (protons or neutrons) emission have been observed in the giant-resonance region due to the one-particle one-hole nature of the charge-exchange excitations \cite{MIKI2017339,PhysRevC.52.604,Zegers2020}. However, the decay by $\gamma$ emission at excitation energies explored here are primarily statistical in nature \cite{Gao2020}. The second assumption is that of the Brink-Axel hypothesis \cite{Brink1955, axel1962}, which states that the $\gamma$SF is independent of the energies, spins, and parities of the initial and final states, and only depends on $E_{\gamma}$. Under these assumptions, the primary matrix $P(E_{\gamma},E_{x})$ can be formulated as \cite{SCHILLER2000,Larsen2011};
\begin{eqnarray}
\ P(E_{\gamma},E_{x}) \propto \rho(E_{x}-E_{\gamma})\mathscr{T}(E_{\gamma}),
\label{eq:one}
\end{eqnarray}   
where $P(E_{\gamma},E_{x})$ is the primary matrix, $\rho(E_{x}-E_{\gamma})$ is the NLD, and $\mathscr{T}(E_{\gamma})$ is the $\gamma$-ray transmission coefficient, which depends only on $E_{\gamma}$. To obtain this primary matrix, an iterative procedure is used as described in Ref. \cite{GUTTORMSEN1987}. The resulting primary matrix $P(E_{\gamma},E_{x})$ is depicted in Fig. \ref{fig:matrices}(d).

For the subsequent analysis of the primary matrix and the extraction of the NLD and $\mathscr{T}(E_{\gamma})$, it is important to exclude regions of the primary matrix that do not meet the requirement that the decay process is statistical. This is achieved by setting lower and upper boundaries in the $E_{x}$ spectrum. It is also necessary to exclude any remaining strongly populated low-lying states from the analysis, as they may bias the iterative process for extracting the first-generation matrix \cite{SCHILLER2000}. This is achieved by setting a lower boundary on $E_{\gamma}$. In practice, these boundaries were varied in the analysis to ensure that the extracted NLD and $\gamma$SF have converged and do not depend on the detailed choice of these boundaries. The boundaries settled in this analysis are $E_{\gamma}^{\text{min}}$=1.6 MeV, $E_{x}^{\text{min}}$=3.2 MeV and $E_{x}^{\text{max}}$=7.2 MeV, as indicated with red lines in Fig. \ref{fig:matrices}(d).

To extract the NLD and $\mathscr{T}(E_{\gamma})$, an iterative $\chi^{2}$-minimization procedure was performed, following Ref. \cite{SCHILLER2000}. The basic idea of this procedure is to construct the theoretical primary matrix by fitting to the experimentally extracted primary matrix. The fit result provides functional forms for both $\rho(E_{x}-E_{\gamma})$ and $\mathscr{T}(E_{\gamma})$. Mathematically, it can be demonstrated that if one solution for the multiplicative functions $\rho(E_{x}-E_{\gamma})$ and $\mathscr{T}(E_{\gamma})$ is known, the fit produces an infinite set of combinations for $\rho(E_{x}-E_{\gamma})$ and $\mathscr{T}(E_{\gamma})$ that reproduce the same $P(E_{\gamma}, E_{x})$ matrix \cite{SCHILLER2000}:
\begin{eqnarray}
\tilde{\rho}(E_{x}-E_{\gamma}) = Ae^{\alpha(E_{x}-E_{\gamma})}\rho(E_{x}-E_{\gamma})
\label{eq:three}
\end{eqnarray}
\begin{eqnarray}
\tilde{\mathscr{T}}(E_{\gamma}) = Be^{\alpha(E_{\gamma})}\mathscr{T}(E_{\gamma}).
\label{eq:four}
\end{eqnarray}
Here, $\rho(E_{x}-E_{\gamma})$ and $\mathscr{T}(E_{\gamma})$ are the solutions, $A$ and $B$ are absolute normalization parameters for the NLD and the $\gamma$-transmission coefficient, respectively, and $\alpha$ denotes the slope parameter. To determine unique values for $A$, $B$, and $\alpha$, a normalization must be performed, which is described in the next section. 

\section{\label{sec:level4}Normalization of the NLD and $\gamma$SF}
\begin{figure}
\hspace*{0.0cm}
\includegraphics[width=0.5\textwidth,clip]{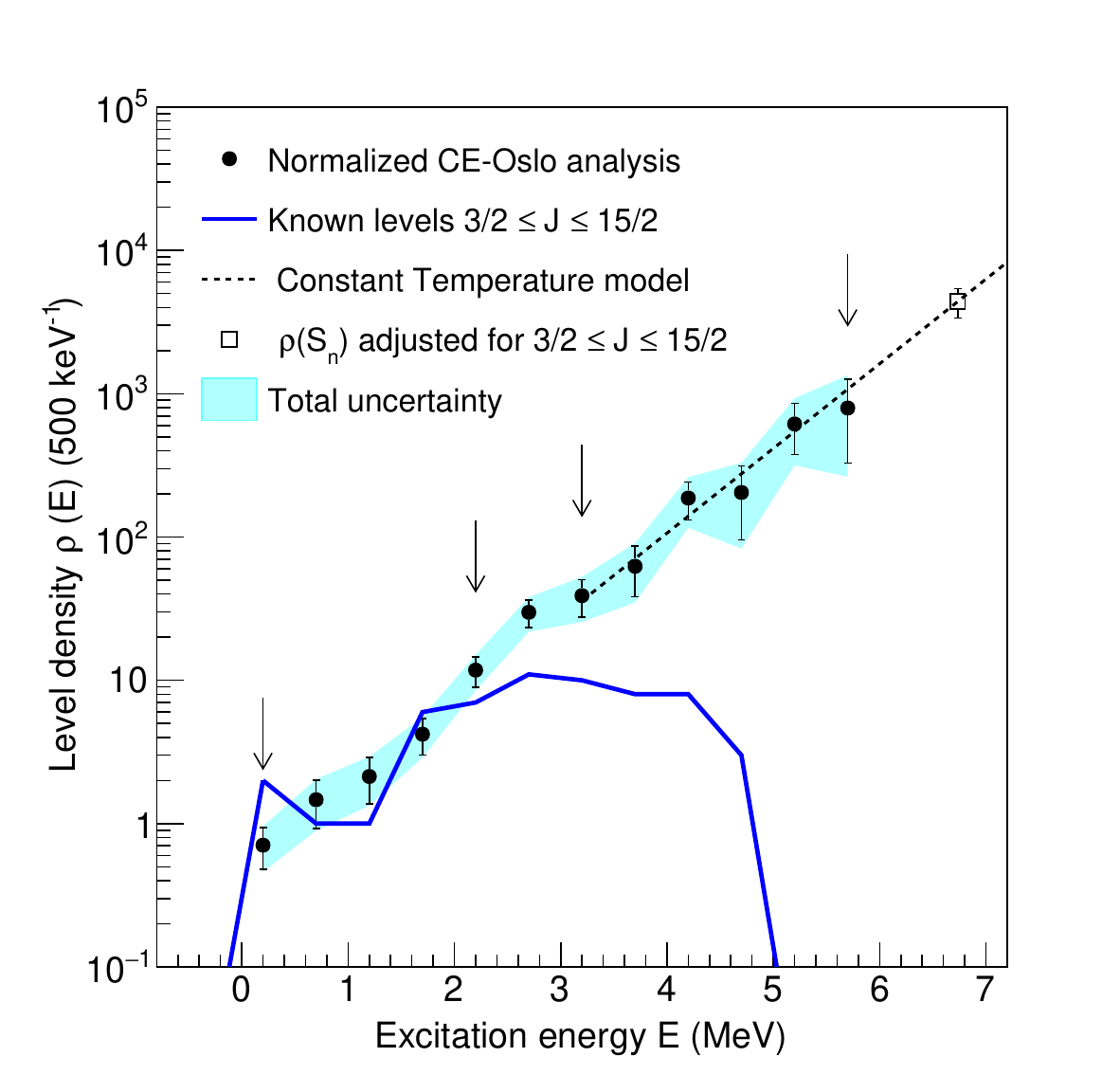}
\caption{The normalized NLD of $^{93}$Zr extracted from the CE-Oslo analysis. Level densities of known discrete levels with spins ranging from $J$ = 3/2 to $J$ = 15/2 are represented by a solid blue line. The lower pair of arrows indicates the region used for normalization. The adjusted NLD for $3/2 \leq J \leq 15/2$ at the $S_{n}$, extracted from neutron resonance data, is indicated with an open square. The NLD fitted to the data using the Constant Temperature model is represented by a dashed black line, with the higher pair of arrows indicating the fitting region. The total uncertainty in the extracted NLD, accounting for both systematic and statistical errors, is illustrated by the cyan shaded area.} 
\label{fig:NLD}
\end{figure}

The normalization of the NLD is necessary to find the values of $A$ and $\alpha$ in Eq. (\ref{eq:three}). At low excitation energies, the extracted level density data points can be constrained to existing data for low-lying states. Based on Ref. \cite{RIPL3}, information about states in $^{93}$Zr is complete up to an excitation energy of 2.4 MeV. 
At the neutron separation energy $S_{n}$, the level density obtained from neutron resonance data is used to constrain $A$ and $\alpha$ in Eq. (\ref{eq:three}).  

In the analysis, it is important to consider the spins of the states that are populated. As discussed in Ref. \cite{Gao2020}, in this experiment the angular momentum transfers induced by the CE reaction are mostly limited to $\Delta L \lesssim 2$. Since $\Delta S=0,1$, and hence $\Delta J \lesssim 3$, states with spins ranging from $J= 3/2$ to $J = 15/2$ can be populated. The decay by $\gamma$ emission will also increase the range of spins of the final levels that can be populated in $^{93}$Zr. Given the uncertainties in the spin distribution of states populated in $^{93}$Zr, two analyses were performed: a primary analysis in which the population of states with spins ranging from $J= 3/2$ to $J = 15/2$ ($\Delta J \leq 3$ from the $^{93}$Nb($9/2^{+}$) ground state)  was assumed, and a secondary analysis in which the population of states with spins ranging from $J= 1/2$ to $J = 17/2$ ($\Delta J \leq 4$ from the $^{93}$Nb($9/2^{+}$) ground state) was assumed. The latter analysis was used to estimate the systematic uncertainties in the extracted quantities due to the uncertainty in the spin distribution. We note that by gating on specific scattering-angle ranges in the ($t$,$^{3}$He) reaction it is possible to constrain the angular momentum transferred, which reduces the systematic uncertainties in the assumed spin distribution. However, in the present data set it was not possible to reduce the angular range investigated because the statistical uncertainties became too large.  

For the normalization of Eq. (\ref{eq:three}) at low excitation energies, one can simply exclude known levels with spins outside of the spin ranges discussed above for the purpose of the normalization procedure. To correct for the number of levels missed at $S_{n}$, one has to use the spin distribution function \cite{Ericson1960}:
\begin{eqnarray}
g(J,\sigma) = \frac{(2J+1)e^{-\frac{(2J+1)^{2}}{2\sigma^{2}}}}{2\sigma^{2}}.
\label{eq:eight}
\end{eqnarray}
In this equation, $\sigma$ is the spin cut-off parameter, for which different models are employed. Here, we employ two commonly used models to estimate the spin cut-off parameter and use the difference between the values as a measure of the systematic uncertainty. 

The first expression for $\sigma$, from Ref. \cite{RMI2009}, is given by: 
\begin{eqnarray}
\sigma^{2}(E_{x}) = 0.0146A^{5/3}\frac{1+\sqrt{1+4a(E_{x}-E_{1})}}{2a},
\label{eq:nine}
\end{eqnarray}
where $a$ and $E_{1}$ represent the level-density parameter and the back-shift parameter, respectively, of the back-shifted Fermi gas (BSFG) model. They were taken from the global systematics of Ref. \cite{RMI2009} and listed in Table \ref{tab:table1}. 

The second approach is taken from Ref. \cite{EB2009}:
\begin{eqnarray}
\sigma^{2}(E_{x}) = 0.391A^{0.675}(E_{x}-0.5Pa^{\prime})^{0.312},
\label{eq:ninea}
\end{eqnarray}
where $A$ is the mass number and $P a^{\prime}$ is the deuteron pairing energy, and listed in Table \ref{tab:table1}. 

In the first approach, the fraction of states missed based on Eq. (\ref{eq:eight}) at $S_{n}$ is 25\% (15\%) for $3/2 \leq J \leq 15/2$ ($1/2 \leq J \leq 17/2$). In the second approach, the fraction of states missed based on Eq. (\ref{eq:eight}) at $S_{n}$ is 16\% (5\%) for $3/2 \leq J \leq 15/2$ ($1/2 \leq J \leq 17/2$).  

The value of the level density at $S_{n}$, $\rho(S_{n})$, is calculated with \cite{SCHILLER2000}:
\begin{eqnarray}
\rho(S_{n}) = \frac{2\sigma^{2}}{D_{0}}\frac{1}{(I_{t}+1)\exp(-\frac{(I_{t}+1)^{2}}{2\sigma^{2}})+I_{t}\exp(-\frac{I_{t}^{2}}{2\sigma^{2}})},
\label{eq:ten}
\end{eqnarray} 
in which $I_{t}$ is the spin of the target, which after neutron capture populates states in $^{93}$Zr, and $D_{0}$ is the average s-wave resonance spacing at $S_{n}$ obtained from Ref. \cite{Tagliente2022}. In the present analysis, the value of $\rho(S_{n})$ is then reduced to account for states with spins outside of the ranges discussed above.

\begin{table}
\caption{\label{tab:table1} Parameter values used in the normalization of NLD and $\gamma$SF}
\begin{ruledtabular}
{\setlength{\extrarowheight}{2pt}%
\begin{tabular}{ll}
Parameter & Value \\
\hline
$S_{n}$ & 6.734 MeV\footnote{From Ref. \cite{Wang_2012}} \\
$D_{0}$ & $(4.0\pm0.5)\times10^{-3}$ MeV \footnote{From Ref. \cite{Tagliente2022}} \\
$a$ & 10.72 MeV$^{-1}$ \footnote{From Ref. \cite{RMI2009}} \\
$E_{1}$ & 0.10 MeV \footnotemark[3] \\
$Pa^{\prime}$ & 0.552 MeV \footnote{From Ref. \cite{EB2009}} \\
$\langle\Gamma_{\gamma0}\rangle$ & $95\pm10$ meV\footnotemark[2] \\
\end{tabular}}
\end{ruledtabular}
\end{table}

In the Oslo method, the NLD is only extracted up to an excitation energy: 
\begin{eqnarray}
E_{x} = (S_{n}-E_{\gamma}^{\text{min}}),
\label{eq:ex}
\end{eqnarray}
due to the exclusion of low-energy $\gamma$ rays in the extraction process \cite{SCHILLER2000}. As shown in Fig. \ref{fig:matrices}(d), $E_{\gamma}^{\text{min}}=1.6$ MeV. However, as discussed above, since the decay by neutron emission only begins at $E_{x}(^{93}$Zr$)\approx 8$ MeV, rather than using $S_{n}$ in Eq. (\ref{eq:ex}), we utilized the value of 7.2 MeV. Therefore, the NLD was extracted up to an excitation energy in $^{93}$Zr of 5.6 MeV. This is more than 1 MeV below $S_{n}$ and for the purpose of normalization with $\rho(S_{n})$, an interpolation of the NLD is required between 5.6 MeV and $S_{n}$. The Constant Temperature model is usually used for this interpolation: \cite{Ericson1959, Ericson1960}:
\begin{eqnarray}
\rho_{CT}(E) = \frac{1}{T_{CT}}\exp\left(\frac{E-E_{0}}{T_{CT}}\right),
\label{eq:eleven}
\end{eqnarray}
where the nuclear temperature ($T_{CT}$) and the energy shift ($E_{0}$) were determined by fitting to the data. The fitted values are included in Table \ref{tab:table2}. 

\begin{table}[b]
\caption{\label{tab:table2} Parameter values obtained in the fitting procedure of NLD and $\gamma$SF. The central values listed are assuming that states with $3/2 \leq J \leq 15/2$ }
\begin{ruledtabular}
{\setlength{\extrarowheight}{2pt}%
\begin{tabular}{ll}
Parameter & Value \\
\hline
$T_{CT}$ &  0.74$^{+0.12}_{-0.09}$ MeV \\
$E_{0}$ & 0.29$^{+0.87}_{-1.18}$ MeV \\
$c_{1}$ &  $(1.1\pm0.6)\times10^{-7}$ MeV$^{-3}$ \\
$c_{2}$ & $0.81\pm0.08$ \\
$c_{3}$ & $1.39\pm0.12$ \\
$\eta$ & $1.1\pm0.3$ (MeV$^{-1}$) \\  
\end{tabular}}
\end{ruledtabular}
\end{table}

The normalized NLD assuming states with spins $3/2 \leq J \leq 15/2$ included is shown in Fig. \ref{fig:NLD}. The lower (higher) set of arrows indicate the regions over which the NLD was fit to the level density of known levels ($\rho(S_{n})$ from neutron resonance data). In the analysis, these fit ranges were adjusted to estimate the systematic uncertainty due to the choice of these ranges. In Fig. \ref{fig:NLD}, the error bars (one standard deviation) on each data point from the Oslo analysis indicate uncertainties in the unfolding and the first-generation method, and the statistical uncertainties, which are the dominant source of uncertainty. The total uncertainties, also include systematic uncertainties due to uncertainties in the fitting caused by the choice of the fitting ranges ($\approx\!15$\%); uncertainties induced by the choice of the spin cut-off parameter model ($\approx\!26$\%); and the uncertainty in $D_{0}$ ($\approx\!10$\%). The same analysis was performed under the assumption that states with spins $1/2 \leq J \leq 17/2$ are included in the analysis, which yielded an NLD that was approximately 15\% higher.        
For the normalization of the $\gamma$SF, it is necessary to determine the value of $B$ in Eq. (\ref{eq:four}) using the average total radiative width at $S_{n}$, $\langle\Gamma_{\gamma0}\rangle$. In principle, $\mathscr{T}(E_{\gamma})$ includes magnetic and electric transitions with any multipolarity. However, it has been shown that dipole transitions $(\Delta L=1)$ dominate the decay in the quasi-continuum region and that the contributions from quadrupole (and higher) multipole transitions are smaller by about two orders of magnitude than that of the E1 contributions \cite{kopecky1990}. Therefore, it was assumed that the contribution originates solely from electric dipole ($E$1) and magnetic dipole ($M$1) transitions, following Ref. \cite{Larsen2011}. Hence, to determine the parameter $B$ in Eq. \eqref{eq:four}, the equation for the average total width associated with s-wave neutron capture $\langle\Gamma_{\gamma0}\rangle$ at $S_{n}$ was used, as detailed in Refs. \cite{kopecky1990,Larsen2011,Voinov2001}:
\begin{eqnarray}
\begin{split}
\langle\Gamma_{\gamma0}\rangle = \langle\Gamma_{\gamma}(S_{n},I_{t}\pm1,\pi_{t})\rangle = \frac{B}{4\pi\rho(S_{n},I_{t}\pm1,\pi_{t})}\\
\times\int_{E_{\gamma}=0}^{S_{n}}dE_{\gamma}\mathscr{T}_{X1}(E_{\gamma})\rho(S_{n}-E_{\gamma})\\
\times\sum_{J=-1}^{1}g(S_{n}-E_{\gamma},I_{t}\pm1/2+J).
\label{eq:fifteen}
\end{split}
\end{eqnarray}
In this equation, $I_{t}$ and $\pi_{t}$ represent the ground-state spin and parity of the target nucleus ($^{92}$Zr) for the neutron-capture reaction, respectively. $\rho(S_{n}-E_{\gamma})$ represents the NLD. The term $1/\rho(S_{n},I_{t}\pm1,\pi_{t})$ corresponds to the $D_{0}$ value. The summation and integration in Eq. (\eqref{eq:fifteen}) were performed over all final states with spins $I_{t} \pm 1/2 + J$ in the spin distribution function. The spin-cutoff parameters in Eqs. \eqref{eq:nine} and \eqref{eq:ninea} were used. The experimental value for $\langle\Gamma_{\gamma0}\rangle$ was obtained from Ref. \cite{Tagliente2022}, and included in Table \ref{tab:table1}. 

Subsequently, the normalized $\gamma$SF, $f(E_{\gamma})$, was calculated using the relationship between $\mathscr{T}(E_{\gamma})$ and $f(E_{\gamma})$ \cite{kopecky1990}:
\begin{eqnarray}
f(E_{\gamma})=\frac{1}{2\pi E_{\gamma}^{3}}B\mathscr{T}(E_{\gamma}) 
\label{eq:sixteen}
\end{eqnarray}

\begin{figure}
\hspace*{0.0cm}
\includegraphics[width=0.5\textwidth,clip]{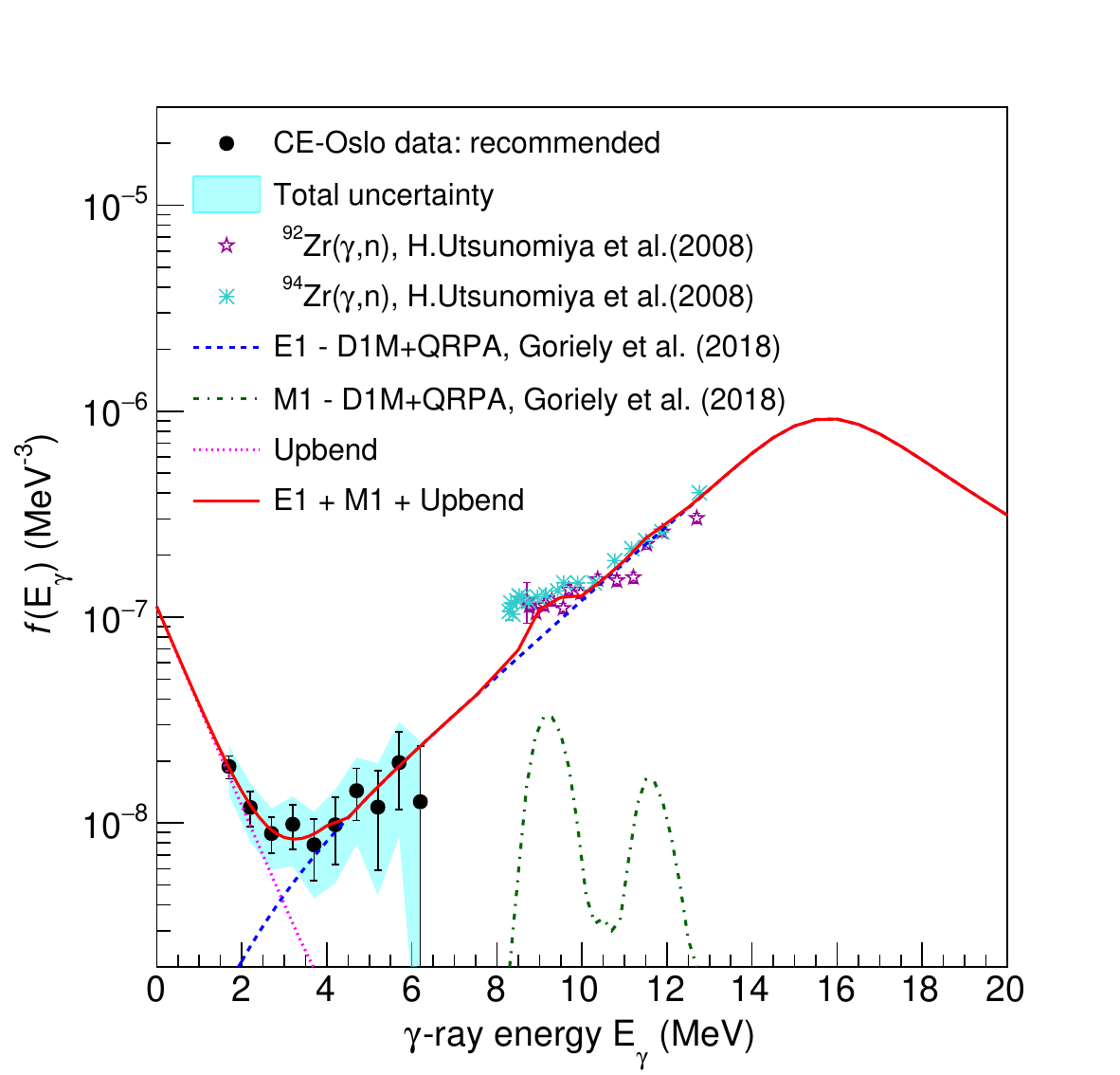}
\caption{The normalized CE-Oslo $f(E_{\gamma})$ of $^{93}$Zr (black solid circles) and the photo-absorption $f(E_{\gamma})$ of $^{92,94}$Zr (blue asterisks and purple stars, respectively) from \cite{Utsunomiya2008} are plotted. The uncertainty band is indicated in cyan. The dashed blue line, dash-dot green line, and the dotted pink line indicate the contributions from $E$1 de-excitations, $M$1 de-excitations, and the upbend, also assumed to be of $M$1 nature. The first two contributions come from the D1M+QRPA model of \cite{goriely2018}.} 
\label{fig:GSF_best_fit}
\end{figure}

The results for $f(E_{\gamma})$ from the Oslo analysis are plotted in Fig. \ref{fig:GSF_best_fit}. The error bars on each data point refer to the statistical uncertainties (one standard deviation) in the Oslo procedure \cite{SCHILLER2000}. The uncertainty band includes the following systematic uncertainties: the uncertainty in $D_{0}$ ($\approx\!10\%$); the uncertainty in $\langle\Gamma_{\gamma0}\rangle$ ($\approx\!10\%$); the error due to the choice of the fitting ranges in the level density normalization ($\approx\!10\%$); the uncertainty induced by the choice of the spin cut-off model ($\approx\!15\%$); and the uncertainty associated with the assumption of the spin range in the Oslo analysis ($\approx\!20\%$). At higher values of $E_{\gamma}$, the statistical uncertainties dominate, but at lower values of $E_{\gamma}$, statistical and systematic uncertainties are comparable.     

\section{\label{sec:level5}Neutron Capture Cross Sections}
The extracted normalized $\gamma$SF, $f(E_{\gamma})$, contains both $E$1 and $M$1 contributions. For the calculation of the $^{92}$Zr($n,\gamma$)$^{93}$Zr cross sections, it is necessary to decompose these contributions and to include $f(E_{\gamma})$ at high values of $E_{\gamma}$. Experimental data for $f(E_{\gamma})$ for $^{93}$Zr are not available for $E_{\gamma}>6$ MeV, including for the Isovector Giant Dipole Resonance (of $E$1 nature), IVGDR, which peaks at $\sim 16$ MeV, and the spin-flip resonance (of $M$1 nature), which peaks at $\sim 9$ MeV. Therefore, $f(E_{\gamma})$ at higher $E_{\gamma}$ was estimated based on data for nearby Zr isotopes by converting experimental photo-absorption cross sections, $\sigma_{\gamma}(E_{\gamma})$, into $f(E_{\gamma})$, using  \cite{RIPL3,Bartholomew1973}:
\begin{eqnarray}
f(E_{\gamma})=\frac{1}{3\pi^{2}\hbar^{2}c^{2}}\frac{\sigma_{\gamma}(E_{\gamma})}{E_{\gamma}}. 
\label{eq:nineteen}
\end{eqnarray}
The data used are from Ref. \cite{Utsunomiya2008}, in which it is shown that the photo-absorption cross sections for the isotopes $^{91,92,94}$Zr are very similar, indicating that using them for the case of $^{93}$Zr is a reasonable approximation. The similarity between the $E$1 and $M$1 resonances at high $E_{\gamma}$ in these nuclei is also supported by the theoretical calculations of Refs. \cite{goriely2016,goriely2018}. The converted $f(E_{\gamma})$ for $^{92,94}$Zr that neighbor $^{93}$Zr based on Ref. \cite{Utsunomiya2008} are shown in Fig. \ref{fig:GSF_best_fit}. In addition to the statistical uncertainties of these data, reported systematic uncertainties \cite{Utsunomiya2008} in the extracted cross section were also considered in the calculation of the total uncertainty band of the extracted $^{92}$Zr($n,\gamma$)$^{93}$Zr cross sections and MACS.

To decompose $f(E_{\gamma})$ into $M$1 and $E$1 components, the data in Fig. \ref{fig:GSF_best_fit} were fitted with the following equation, in which $c_{1,2,3}$ and $\eta$ are the fitted parameters:
\begin{eqnarray}
\begin{split}
f(E_{\gamma}) = c_{1}e^{-\eta E_{\gamma}}+c_{2}f_{E1}^{D1M+QRPA}(E_{\gamma})\\
+c_{3}f_{M1}^{D1M+QRPA}(E_{\gamma}).
\label{eq:eighteen}
\end{split}
\end{eqnarray}  
In this equation, $f_{E1}^{D1M+QRPA}(E_{\gamma})$ and $f_{M1}^{D1M+QRPA}(E_{\gamma})$ are theoretical calculations for the IVGDR and spin-flip $M$1 resonance in $^{93}$Zr, respectively, using the Hartree-Fock-Bogoliubov (HFB) plus Quasiparticle Random-Phase Approximation (QRPA) method based on the Gogny D1M interaction, as detailed in Ref. \cite{goriely2018} and available in the database for photon-strength functions in TALYS \cite{koning2023}. This model was chosen based on previous studies of $^{91}$Zr and $^{92}$Zr \cite{9192Zr}, where it was used to benchmark the measured $^{90,91}$Zr($n,\gamma$) cross sections and MACS. 
The inclusion of an $M1$ component is motivated by previous experimental studies \cite{Utsunomiya2008, Iwamoto2012, 9192Zr}, which identified $M1$ resonance contributions in the isotopes $^{90,91,92,94}$Zr at centroid energies of about  9 MeV. Based on these findings, the centroid energy of the $M$1 resonance in the D1M+QRPA calculation was shifted to this energy prior to the fitting. In the calculation of the neutron-capture cross sections described below, it was found that while the inclusion of the $M$1 resonance does not significantly impact the estimated cross section, its inclusion improves the overall fit of $f(E_{\gamma})$ and reduces the uncertainty in the normalization of the $E1$ component. To account for uncertainties due to incomplete knowledge of the $M$1 contribution in the fit, an additional systematic uncertainty in the normalization of the $E$1 component of 5\% was included in the error analysis of the estimated neutron-capture cross sections.  
      
The term $e^{-\eta E_{\gamma}}$ represents the $M$1 upbend (also referred to as low-energy enhancement (LEE)) as described in \cite{goriely2018}, which is necessary to explain the observed enhancement of the $\gamma$SF below approximately 2.5 MeV. Refs. \cite{Schwengner2013, sieja2017, brown2014, karampagia2017} show that the enhancement occurs near closed shells. In Ref. \cite{Schwengner2013}, it was demonstrated that this phenomenon is commonly found in nuclei with $Z \approx$ 40 and $N \approx$ 50.  It was assumed that there is no $E$1 contribution to the upbend, as suggested by the shell-model calculations of Ref. \cite{sieja2017}. However, we note that this assumption does not affect the extracted neutron-capture cross sections. 

The fit results are included in Fig. \ref{fig:GSF_best_fit} and Table \ref{tab:table2}. The fit values of $c_{2}$ and $c_{3}$ indicate that minor scalings had to be applied to the D1M+QRPA theoretical calculations based on Ref. \cite{goriely2018} to obtain the best fit.     
 
\begin{figure}
\hspace*{0.0cm}
\includegraphics[width=0.5\textwidth,clip]{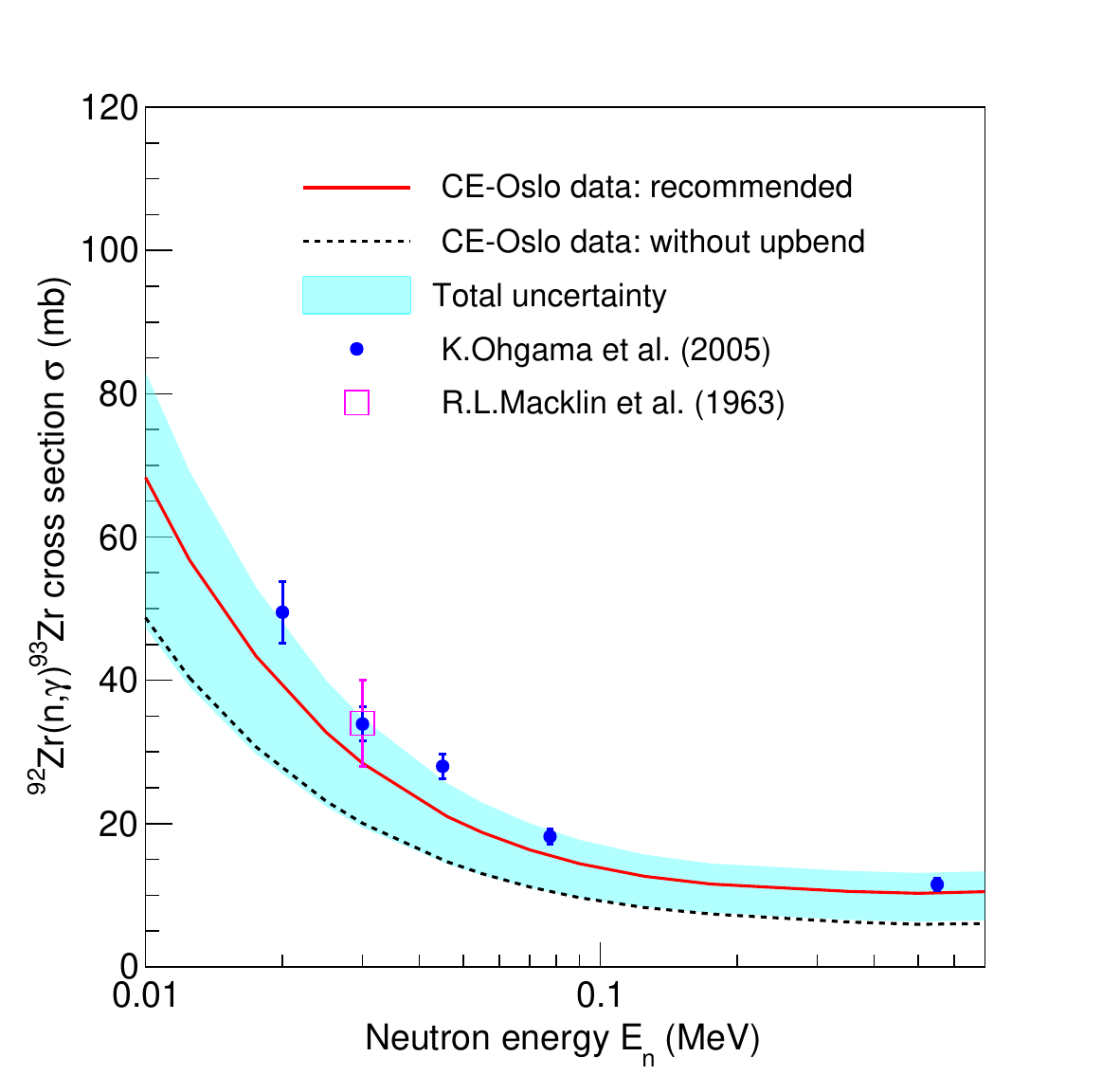}
\caption{The extracted CE-Oslo $^{92}$Zr($n,\gamma$)$^{93}$Zr cross sections are shown in a red solid line. The directly measured $^{92}$Zr$(n,\gamma)^{93}$Zr cross sections, obtained from \cite{OHGAMA2005,Macklin1963}, are included. The total uncertainty, which includes both the systematic and statistical uncertainties, is depicted by the cyan shaded area. The effect of removing the contribution from the upbend component in $f(E_{\gamma})$ to the $^{92}$Zr$(n,\gamma)^{93}$Zr cross sections  is shown as a black dashed line. }
\label{fig:cross_sections}
\end{figure}

The experimentally extracted NLD and decomposed $f(E_{\gamma})$ for $^{93}$Zr, were utilized as inputs for calculating the $^{92}$Zr($n,\gamma$)$^{93}$Zr cross sections. The TALYS 1.96 package \cite{koning2023} was used for performing the calculations using the Hauser-Feshbach theoretical framework, as described in Ref. \cite{RAUSCHER20001}. The optical model potential parameters from Koning and Delaroche \cite{KONING2003} were used to detail the interaction between the neutron and $^{92}$Zr. The resulting $^{92}$Zr($n,\gamma$)$^{93}$Zr cross sections for neutron energies $0.01<E_{n}<0.7$ MeV are shown (red solid line) in Fig. \ref{fig:cross_sections}, together with the 1$\sigma$ uncertainty band. A second calculation was performed in which the contribution from the upbend was switched off, as indicated by the dashed black line, which reduces the neutron-capture cross section by 30-43\%, depending on the value of $E_{n}$. 

The experimental data from Refs. \cite{OHGAMA2005, Macklin1963} are also shown in Fig. \ref{fig:cross_sections}. The indirectly constrained $^{92}$Zr($n,\gamma$)$^{93}$Zr cross sections are slightly lower, but consistent with the directly measured cross sections within uncertainties. The inclusion of the upbend, which was observed in the $^{93}$Nb($t$,$^{3}$He$+\gamma$) data, is important to achieve consistency, as the neutron-capture cross section without this contribution is significantly below the measured cross sections otherwise. It is noted that there is no experimental data for the $\gamma$-ray strength function in the region $6<E_{\gamma}<8$ MeV. The presence of $E1$ pygmy dipole strength in this region could increase the neutron-capture cross section. However, no significant pygmy dipole strength was observed in this region in other Zr isotopes \cite{9192Zr}.

\begin{figure}
\hspace*{0.0cm}
\includegraphics[width=0.5\textwidth,clip]{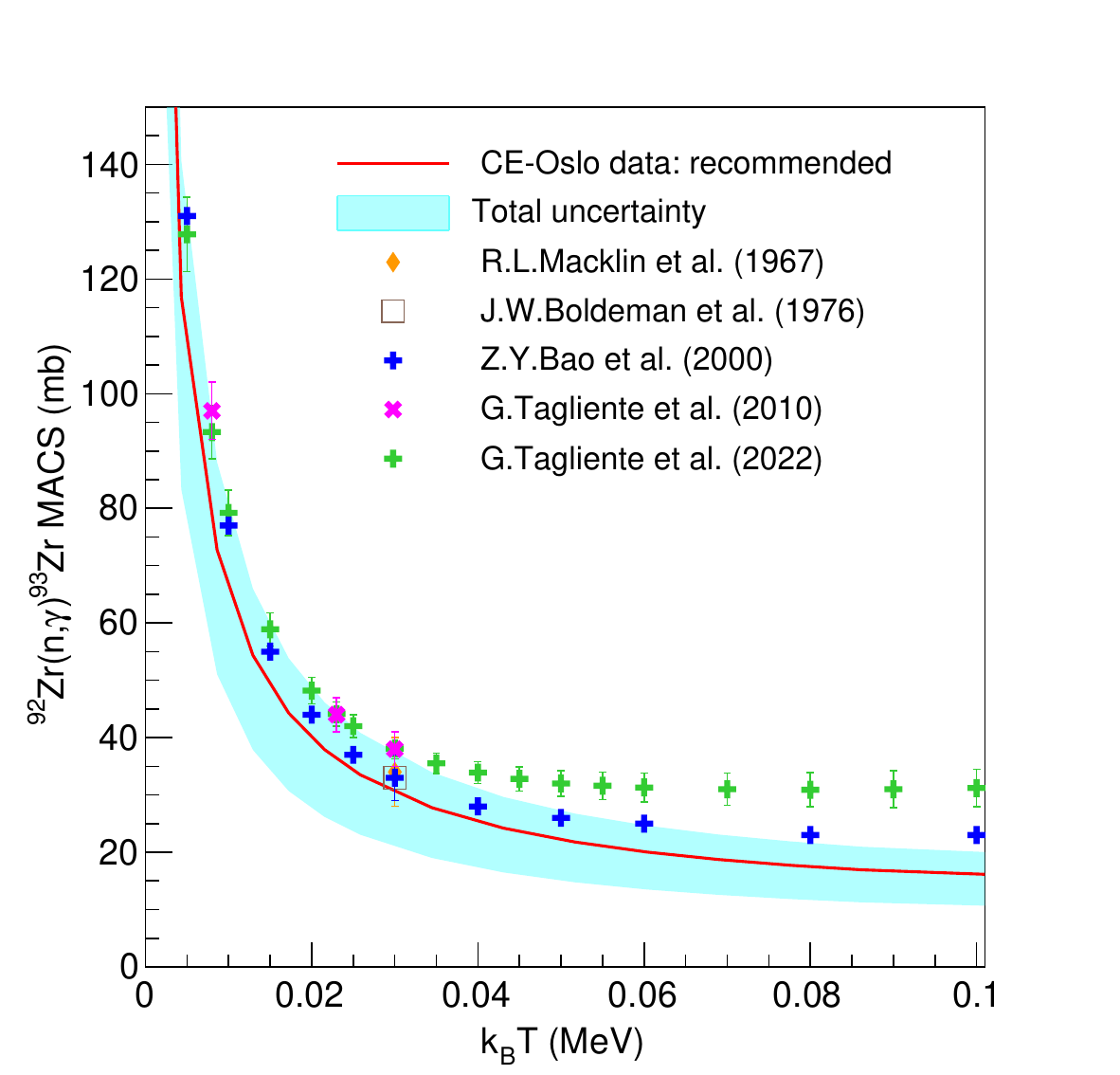}
\caption{The calculated CE-Oslo $^{92}$Zr($n,\gamma$)$^{93}$Zr Maxwellian averaged cross sections (MACS) are shown in a red solid line. The calculated $^{92}$Zr$(n,\gamma)^{93}$Zr Maxwellian-averaged cross sections, obtained from \cite{macklin1967,BOLDEMAN1976,Bao2000,Tagliente2010,Tagliente2022}, are included. The total uncertainty, which includes both the systematic and statistical uncertainties, is depicted by the cyan shaded area.} 
\label{fig:cross_sections_maxwellian}
\end{figure}

Using the indirectly extracted $^{92}$Zr($n,\gamma$)$^{93}$Zr cross sections, the MACS was also calculated with TALYS \cite{koning2023}. The result for $0<k_{B}T<0.1$ MeV is shown in Fig. \ref{fig:cross_sections_maxwellian}, together with the 1$\sigma$ uncertainty band. Figure \ref{fig:cross_sections_maxwellian} also includes the MACS based on experiment neutron-resonance data from Refs. \cite{macklin1967,BOLDEMAN1976,Bao2000,Tagliente2010,Tagliente2022}. Following the above results for the $^{92}$Zr($n,\gamma$)$^{93}$Zr cross sections, the estimated MACS based on the indirect information are slightly lower than the MACS based on the neutron-resonance data, but consistent within the uncertainties for $k_{B}T$ below approximately 30 keV. With increasing temperature, there are signficant differences between the tabulation of Ref. \cite{Bao2000} and Ref. \cite{Tagliente2022}. At temperatures above $k_{B}T\approx30$ keV, neutron-capture cross sections for $E_{n}\gtrsim81$ keV become increasingly important \cite{Tagliente2022}, but neutron-resonance data are not available. Therefore, in Ref. \cite{Tagliente2022}, $^{92}$Zr($n,\gamma$)$^{93}$ cross sections recommended in ENDF/B-VIII.0 \cite{BROWN20181}, JEFF-3.3 \cite{plompen_cabellos_saint}, and JENDL-4.0 \cite{SHIBATA01012011} were used. By using the NLD and $f(E_{\gamma})$ obtained in this work, the MACS at higher temperatures are obtained based on the combined analysis of experimental data and theoretical models. As shown in Fig. \ref{fig:cross_sections}, our data reproduce the only directly measured experimental neutron-capture cross section for $E_{n}>100$ keV available. We find that the MACS at $k_{B}T=100$ keV estimated based on the current work are about 50\% lower than the value estimated in Ref. \cite{Tagliente2022} and 30\% lower than the values from \cite{Bao2000}, although the present results are within uncertainties consistent with the latter. In Ref. \cite{Tagliente2022}, it was concluded that: ``improved capture cross section data are required for a new evaluation of the $^{92}$Zr($n,\gamma$)$^{93}$Zr cross section in the energy region above 80 keV''. The present results provide such information through indirect techniques. 

\section{\label{sec:level6}Summary and Outlook}
In this work, the charge-exchange Oslo method was successfully introduced using the $^{93}$Nb($t$,$^{3}$He$+\gamma$) reaction at 115 MeV/u. The excitation energy in $^{93}$Zr was determined via a missing-mass calculation using the momentum of the $^{3}$He ejectile measured in the S800 Spectrograph, while the $\gamma$ rays from the deexcitation of $^{93}$Zr were detected with GRETINA. The matrix of $\gamma$ energy versus excitation energy was analyzed using the Oslo method, and the nuclear level density and $\gamma$-ray strength function were extracted. The latter was decomposed into its contributions from electric and magnetic dipole strengths, including a low--$E_{\gamma}$ $M$1 upbend. Subsequently, the neutron-capture cross section on $^{92}$Zr was estimated and compared with results from direct measurements. Bare neutron-capture cross sections and Maxwellian averaged cross sections (MACS) were compared. In the region $E_{n}<81$ keV, where neutron resonance data are available, a good consistency between the direct measurements and the indirect results from the charge-exchange Oslo analysis was found. The contribution from the upbend in the extracted $\gamma$-ray strength function is important to achieve the consistency. 
For higher neutron energies, previous estimates relied on database compilations to estimate the capture cross section. In the present work, the results from the Oslo analysis, in combination with auxiliary experimental data and theoretical estimates for the $\gamma$-ray strength function, suggest that the Maxwellian averaged cross section at high temperatures is lower by 50\% (30\%) than the estimates from Refs. \cite{Tagliente2022} and  \cite{Bao2000,Tagliente2022}, respectively, although the results are within uncertainties consistent with the tabulation of  \cite{Bao2000}.   

The successful development of the charge-exchange Oslo method paves the way for further studies focused on nuclear astrophysics and applications for which ($n$,$\gamma$) reactions are important. Charge-exchange reactions at intermediate energies are the preferred probe for extracting Gamow-Teller transition strengths of interest for nuclear astrophysics, especially when direct studies via electron-capture or $\beta$-decay experiments are not feasible. Therefore, charge-exchange reactions with $\gamma$-ray coincidence make it possible to deduce both ($n$,$\gamma$) cross sections and weak-interaction rates in a single measurement. The longer-term goal is to extend these capabilities to ($p$,$n+\gamma$) reactions in inverse kinematics with rare isotope beams.

\section{\label{sec:level7}Acknowledgements}

We thank the NSCL staff for their support during the preparations for and conducting of the experiment. This work was supported by the US National Science Foundation (NSF) under PHY-2209429, PHY-1913554, PHY-1811855, Cooperative Agreement PHY-156554 (NSCL), PHY-1430152 (JINA Center for the Evolution of the Elements) and PHY-1927130 (AccelNet-WOU: International Research Network for Nuclear Astrophysics [IReNA]). GRETINA was funded by the US Department of Energy, in the Office of Nuclear Physics of the Office of Science under award DE-SC0023633 (MSU). Operation of the array at NSCL was supported by DOE under Grants No. DE-SC0014537 (NSCL) and No. DE-AC02-05CH11231 (LBNL). B.G. thanks the support by the China Scholarship Council as part of the FRIB-CSC Fellowship, and the National Key Research and Development program (MOST 2022YFA1602304). J.C.Z. thanks the support by Funda\c{c}$\tilde{a}$o de Amparo a Pesquisa do Estado de S$\tilde{a}$o (FAPESP) under Grant No. 2018/04965-4. A.C.L. thanks the funding from the Research Council of Norway under Project No. 316116, and support from the Norwegian Nuclear Research Center under Project No. 341985.

\bibliography{biblio}

\end{document}